\documentclass[a4paper,11pt]{article}
\usepackage{pos}

\newcommand{\GE}{G_{\rm E}}
\newcommand{\GM}{G_{\rm M}}

\newcommand{\eV}[1]{\epsilon^{\rm mid}_{{\rm E},{#1}}}
\newcommand{\eM}{\epsilon^{\rm mid}_{\rm M}}
\newcommand{\GEmid}{G^{\rm mid}_{\rm E,4}}
\newcommand{\GMmid}{G^{\rm mid}_{\rm M}}

\title{Nucleon-pion-state contamination in lattice computations of the nucleon electromagnetic form factors}
\ShortTitle{$N\pi$-state contamination in nucleon electromagnetic form factor calculations} 

\author*[a]{Oliver B{\"a}r}
\author[a]{Haris \v{C}oli\'{c}}

\affiliation[a]{Humboldt Universität zu Berlin,\\
  Newtonstrasse 15, 12489 Berlin, Germany}

\emailAdd{obaer@physik.hu-berlin.de}
\emailAdd{haris.colic@student.hu-berlin.de}

\abstract{The nucleon-pion-state contributions to QCD two-point and three-point functions relevant for lattice calculations of the nucleon electromagnetic form factors are studied in chiral perturbation theory. 
To leading order the results depend on a few experimentally known low-energy constants only, and the nucleon-pion-state contribution to the form factors can be estimated. The nucleon-pion-state contribution to the electric form factor $G_{\rm E}(Q^2)$ is at the +5 percent level for a source-sink separation of 2 fm, and it increases with increasing momentum transfer $Q^2$. For the magnetic form factor  the nucleon-pion-state contribution leads to an  underestimation of $G_{\rm M}(Q^2)$ by about $-5$ percent that decreases with increasing $Q^2$. 
For smaller source-sink separations that are accessible in present-day lattice simulations the impact is larger. Although  the ChPT results may not be applicable for these time separations a comparison with recent lattice data works reasonably well.}

\FullConference{%
 The 38th International Symposium on Lattice Field Theory, LATTICE2021
  26th-30th July, 2021
  Zoom/Gather@Massachusetts Institute of Technology
}

\begin{document}
\maketitle

\section{Introduction}
While physical point simulations eliminate the need for the chiral extrapolation and the systematic error associated with it, this advantage comes at a prize. Not only are such simulations numerically demanding, two issues typically get worse the smaller the pion mass is in a numerical simulation: Firstly, the signal-to-noise problem \cite{Parisi:1983ae,Lepage:1989hd} limits the accessible time separations in correlation functions that are computed to obtain the observables of interest. Secondly, the contamination in these correlation functions due to multi-particle states with additional pions increases rapidly the smaller the pion mass is.

It has been shown that chiral perturbation theory (ChPT) \cite{Weinberg:1978kz,Gasser:1983yg,Gasser:1984gg} provides a very useful tool to study the second issue \cite{Tiburzi:2009zp,Bar:2012ce}. The impact of the dominant two-particle nucleon-pion ($N\pi$) state-contamination in lattice calculations of various nucleon observables has been investigated, for instance in the nucleon mass \cite{Bar:2015zwa}, various nucleon charges \cite{Bar:2016uoj}, moments of parton distribution functions \cite{Bar:2016jof}, and in the axial and pseudoscalar form factors of the nucleon \cite{Bar:2018xyi,Bar:2019gfx}. Especially for the form factors the ChPT results provide a lot of understanding for the excited-state contamination, in particular for resolving the so-called "PCAC puzzle" \cite{Bali:2018qus}: A low-energetic $N\pi$ excited-state has been identified to be the culprit for the  apparent violation of the generalized Goldberger-Treiman relation \cite{Bar:2018xyi,Bar:2019igf,Bali:2019yiy,Jang:2019vkm}.

In this contribution we report our results for the $N\pi$-contamination in the electromagnetic form factors of the nucleon. Here we focus on the main results, additional details about the calculation can be found in Ref.\ \cite{Bar:2021crj}. For a  general overview of the method see also the review articles \cite{Bar:2017kxh,Bar:2017gqh}.

\section{Electromagnetic form factors}

Assuming isospin symmetry the matrix element of the isovector vector current  between two single nucleon states can be decomposed into the Dirac and Pauli form factors , $F_1(q^2)$ and $F_2(q^2)$, respectively. These are related to the electric and magnetic Sachs form factors $G_{\rm E}(q^2)$ and $G_{\rm M}(q^2)$ according to 
\begin{eqnarray}
\GE(q^2)=F_1(q^2) + \frac{q^2}{4M_N^2}F_2(q^2)\,,\label{SachsFFGE}\\
\GM(q^2)=F_1(q^2)+F_2(q^2)\,.\label{SachsFFGM}
\end{eqnarray}
The method to compute the Sachs form factors with lattice QCD simulations is well-established, see Refs.\ \cite{Alexandrou:2018sjm,Jang:2019jkn,Djukanovic:2021cgp,Park:2021ypf} for some recent calculations. We skip the technical details  and refer to these Refs.\ and \cite{Bar:2021crj} for the relevant formulae of what we are describing in words here. 

In short, one computes the nucleon 2-point (pt) and 3-pt functions involving the vector current $V_{\mu}$. The interpolating nucleon fields are separated by the source-sink separation $t$, and the vector current  is placed at operator insertion time $t'$  between source and sink. An appropriate generalized ratio $R_{\mu}(\vec{q},t,t')$ of the 2-pt and 3-pt functions is formed that, by construction,  converges to constant values $\Pi_{\mu}(\vec{q})$ when all time separations go to infinity. The $\Pi_{\mu}(\vec{q})$ are proportional to the form factors we are interested in. 

In practice, however, one is limited to finite time separations with the source-sink separation well below 2 fm in present-day simulations. Consequently, following the procedure described before we obtain {\em effective form factors} that contain a non-vanishing excited-state contamination. Quite generally, the effective form factors can be written as ($Q^2=-q^2$)
\begin{eqnarray}
G^{\rm eff}_{\rm X}(Q^2,t,t')\, = \,G_{\rm X}(Q^2)\bigg[ 1 + \epsilon_{\rm X}(Q^2,t,t')\bigg],\quad X\,=\,E,M\,,\label{EffGX}
\end{eqnarray}
with $G_{\rm X}(Q^2)$ being the actual form factor we are interested in, and an excited-state contribution $ \epsilon_{\rm X}(Q^2,t,t')$ that vanishes for $t,t',t-t'\rightarrow \infty$. 

Based on the effective form factors we can define the plateau or midpoint estimates for the form factors, the latter being defined as
\begin{eqnarray}
G_{\rm X}^{\rm mid}(Q^2,t) & = & G_{\rm X}^{\rm eff}(Q^2,t,t'=t/2) \,.
\end{eqnarray}
It only depends on the squared momentum transfer $Q^2$ and the source-sink separation $t$. 
In practice, the midpoint and the plateau estimates are essentially the same, at least for the small momentum transfers we are interested in in the following.

\section{The $N\pi$ excited-state contamination in ChPT}

%
\begin{figure}[t]
\begin{center}
\includegraphics[scale=0.9]{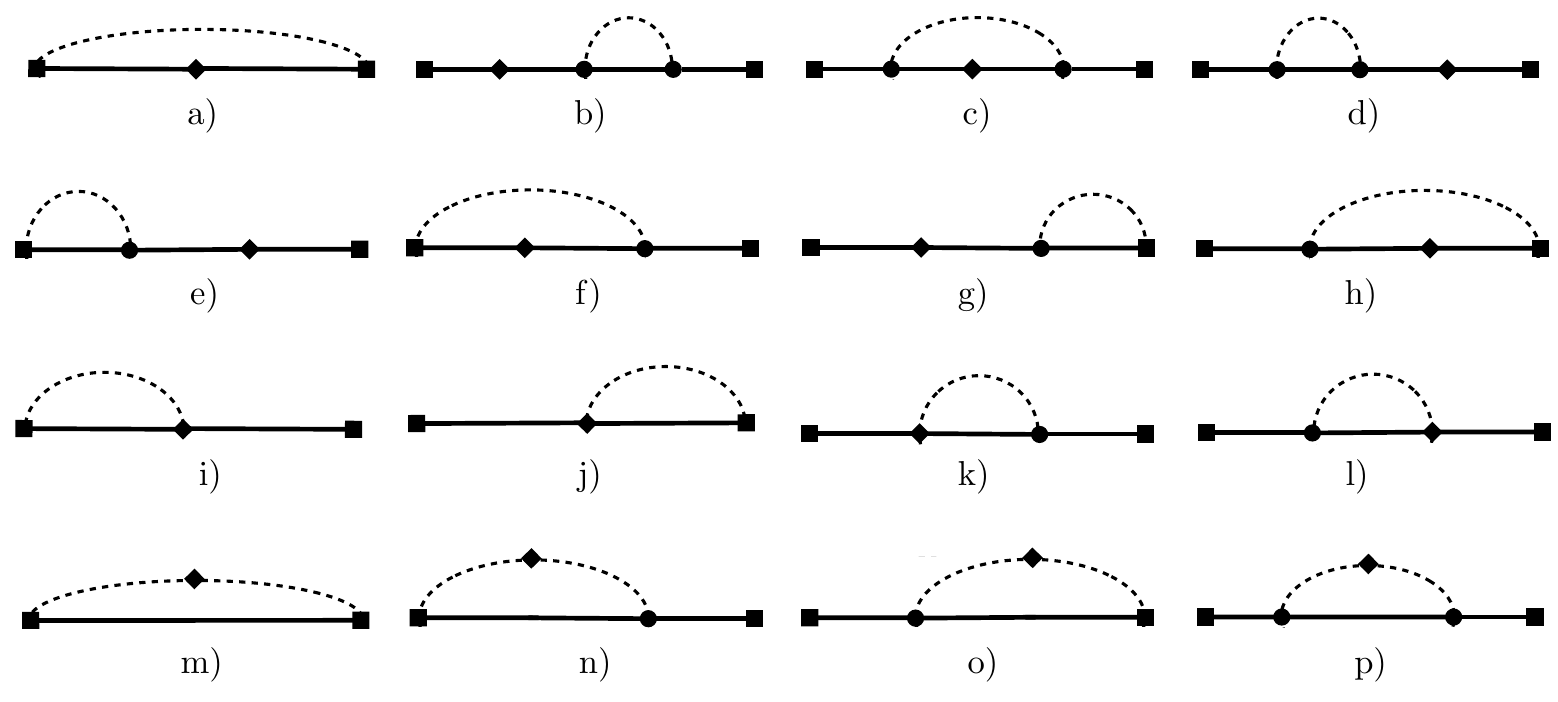}\\[0.3ex]
\caption{Feynman diagrams for the leading $N\pi$ contribution in the vector current 3-pt function. 
Squares represent the nucleon interpolating fields at times $t$ and $0$, and the diamond stands for the vector current at insertion time $t'$.
The solid  and dashed lines represent a nucleon and pion propagator, respectively. }
\label{fig:Npidiags3pt}
\end{center}
\end{figure}

The procedure we have described in the previous section is not only followed in numerical QCD simulations, the same steps can be taken in Baryon ChPT \cite{Gasser:1987rb}. Doing so we obtain the part $\epsilon_{\rm X}^{N\pi}(Q^2,t,t')$ of the excited-state contribution that is caused by $N\pi$ states. This part is expected to be the dominant one for large time separations, for the energy gap to the single nucleon ground state is smaller for low-energetic $N\pi$ states than for other excited states with the same quantum numbers as the nucleon.

In ChPT the relevant correlation functions are computed perturbatively. Figure \ref{fig:Npidiags3pt} shows the sixteen 1-loop Feynman diagrams that contribute to the leading order (LO) result for the $N\pi$ contribution in the 3-pt function with the vector current \cite{Bar:2021crj}. In addition, there are four more diagrams for the 2-pt function \cite{Bar:2015zwa}. Following the procedure described above one obtains analytic but cumbersome results for $\epsilon_{\rm X}^{N\pi}(Q^2,t,t')$, see Ref.\ \cite{Bar:2021crj}. The results depend on
three low-energy-coefficients (LECs), and to the order we are working these can be fixed by setting them to their experimentally known values: The pion decay constant $f_{\pi}= 93$ MeV, the axial charge $g_A=1.27$  and the difference of the magnetic moments of the proton and neutron, $\Delta \mu = \mu_p - \mu_n =4.706$  \cite{Tanabashi:2018oca}. In addition, the pion and nucleon masses are set to their approximate physical values, $M_{\pi}=140$ MeV and $M_N=940$ MeV. 

We assume a finite spatial volume with extent $L$ and periodic boundary conditions in each spatial direction. The size of the volume is determined by fixing the value for the dimensionless number $M_{\pi}L$.
We emphasize that the LO results do not depend on any (unknown) LECs associated with the nucleon interpolating fields. This makes the ChPT results predictive and useful. At higher orders in the chiral expansion this property will be lost.

%
\begin{figure}[t]
\begin{center}
\includegraphics[scale=1.0]{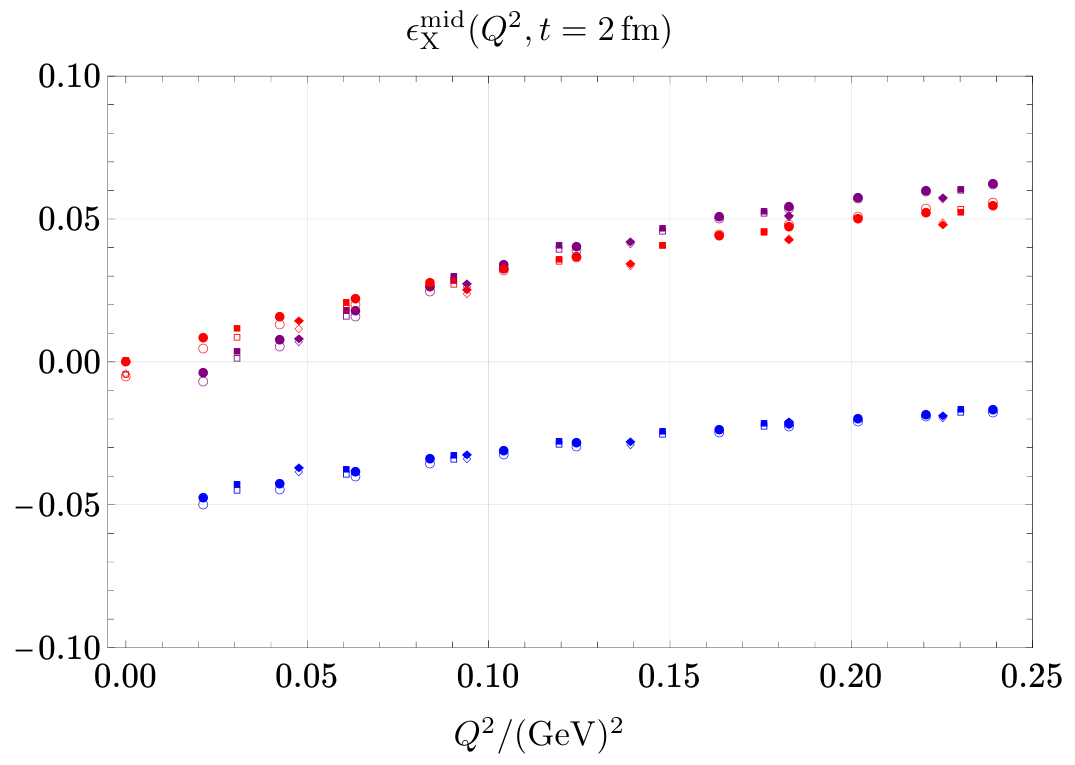}\\[0.3ex]
\caption{The relative deviation $\epsilon^{\rm mid}_{\rm X}(Q^2,t=2\,{\rm fm})$ for the midpoint estimates as a function of $Q^2$. Results for $X=E,4$ in red, $X=M$ in blue and $X=E,2$ in purple. Results for three different spatial volumes with $M_{\pi}L=4$ (diamonds), 5 (squares) and 6 (circles). Open symbols stand for an approximation with the excited-to-excited state contribution being ignored, see Ref.\ \cite{Bar:2021crj}.}
\label{fig:epsMid}
\end{center}
\end{figure}

Figure \ref{fig:epsMid} shows the $N\pi$ contamination for the midpoint estimates, i.e.
\begin{eqnarray}
\epsilon^{\rm mid}_{\rm X}(Q^2,t)=\epsilon^{\rm eff}_{\rm X}(Q^2,t,t'=t/2),
\end{eqnarray}
 as a function of $Q^2$ for a source-sink separation $t=2\,$fm. Results are shown for three different spatial volumes with $M_{\pi}L=4$ (diamonds), 5 (squares) and 6 (circles). The results for a given volume show a smooth $Q^2$ dependence. A small FV effect is visible when we compare the results for $M_{\pi}L=4$ and $6$. However, it is much smaller than the anticipated precision of the LO results.

The electric form factor can be extracted using the vector current 3-pt function with either the time-like component $\mu=4$ or a spatial component $\mu=1,2,3$. For the former case 
$\eV{4}$ is positive and rises monotonically to about $+5\%$ for $Q^2=0.25\,{\rm GeV}^2$. It vanishes exactly for $Q^2=0$ as a result of the Ward identity stemming from vector current conservation if isospin is conserved. For the second case the deviation $\eV{2}$ is close to $\eV{4}$. The difference between the two increases for small $Q^2$. Still, the difference is not pronounced enough to clearly favor one choice over the other. The deviation $\eM$ for the magnetic form factor is negative and ranges between $-5\%$ and $-2\%$ for the momenta displayed in the figure. Here, in contrast to $\eV{4}$, the deviation increases for $Q^2$ getting smaller. 

We emphasize that the excited-state contamination shown in figure \ref{fig:epsMid} is the cumulative contribution of many $N\pi$ states with increasing spatial momenta. The exact number depends on the size of the spatial volume, since this determines the allowed discrete spatial momenta of both pion and nucleon, and subsequently the energy of the 2-particle $N\pi$ state. For $M_{\pi}L=4$ the deviation $\epsilon^{\rm mid}_{\rm X}(Q^2,t)$ in figure \ref{fig:epsMid} is the contribution of five $N\pi$ states with the five smallest non-vanishing pion momenta, and this number increases to twelve for $M_{\pi}L=6$.

%
\begin{figure}[t]
\begin{center}
\includegraphics[scale=1.0]{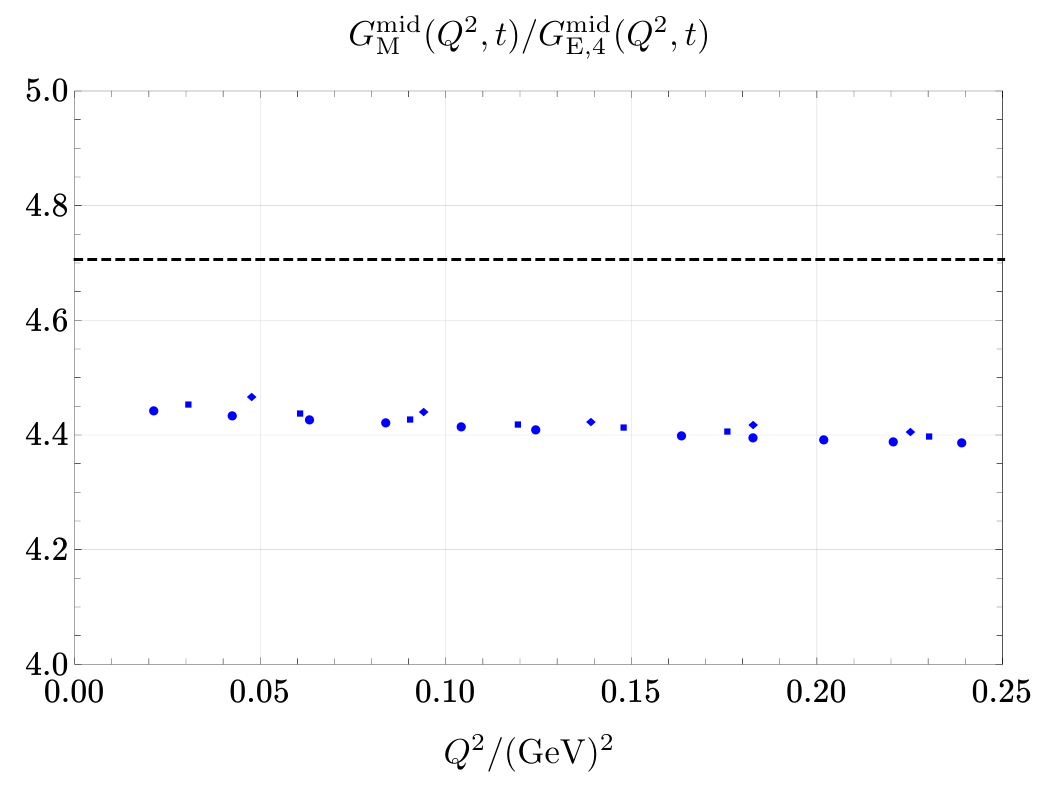}\\[0.3ex]
\caption{\label{figGMmidOverGEmid} The ratio $\GMmid(Q^2,t)/\GEmid(Q^2,t)$ for $t=2$ fm and $M_{\pi}L=4$ (diamonds), 5 (squares) and 6 (circles). The dashed line shows the value $\mu_{p}-\mu_{n}=4.71$. 
}
\end{center}
\end{figure}

Fig.\ \ref{figGMmidOverGEmid} shows the ratio $\GMmid(Q^2,t)/\GEmid(Q^2,t)$ as a function of $Q^2$, again for $t=2$ fm and various $M_\pi L$ values. To a very good approximation this ratio is constant, it varies by less than 2 percent over the range of $Q^2$ displayed in the figure. The mild $Q^2$ dependence is anticipated since the slopes of $\eV{4}(Q^2,t)$ and $\eM(Q^2,t)$ are similar and essentially cancel in the ratio. However, this flat $Q^2$ behavior should not be misinterpreted as the absence of the $N\pi$-state contamination: The ratio is about 6\% below $\mu_{p}-\mu_{n}$, the value it assumes at vanishing momentum transfer without the $N\pi$ contamination.

\section{Comparison with lattice data}

For figure \ref{fig:epsMid} a source-sink separation of 2 fm was chosen. This is sufficiently large to expect ChPT to give reliable results for the $N\pi$ contamination \cite{Bar:2016uoj}. In present-day simulations the accessible source-sink separations are substantially smaller, typically not larger than 1.5 fm. 
For example, recent simulations by the PACS collaboration \cite{Ishikawa:2018rew}  were performed with a source-sink simulation $t \approx 1.3$ fm, which is fairly small. Still, the simulations were performed with an almost physical pion mass, $M_{\pi} \approx 146$ MeV, and a fairly large volume satisfying $M_\pi L\approx 6.0$. This implies a smallest non-vanishing momentum transfer of about $Q^2 \approx 0.024\,{\rm GeV}^2$. 

%
\begin{figure}[t]
\begin{center}
\includegraphics[scale=1.0]{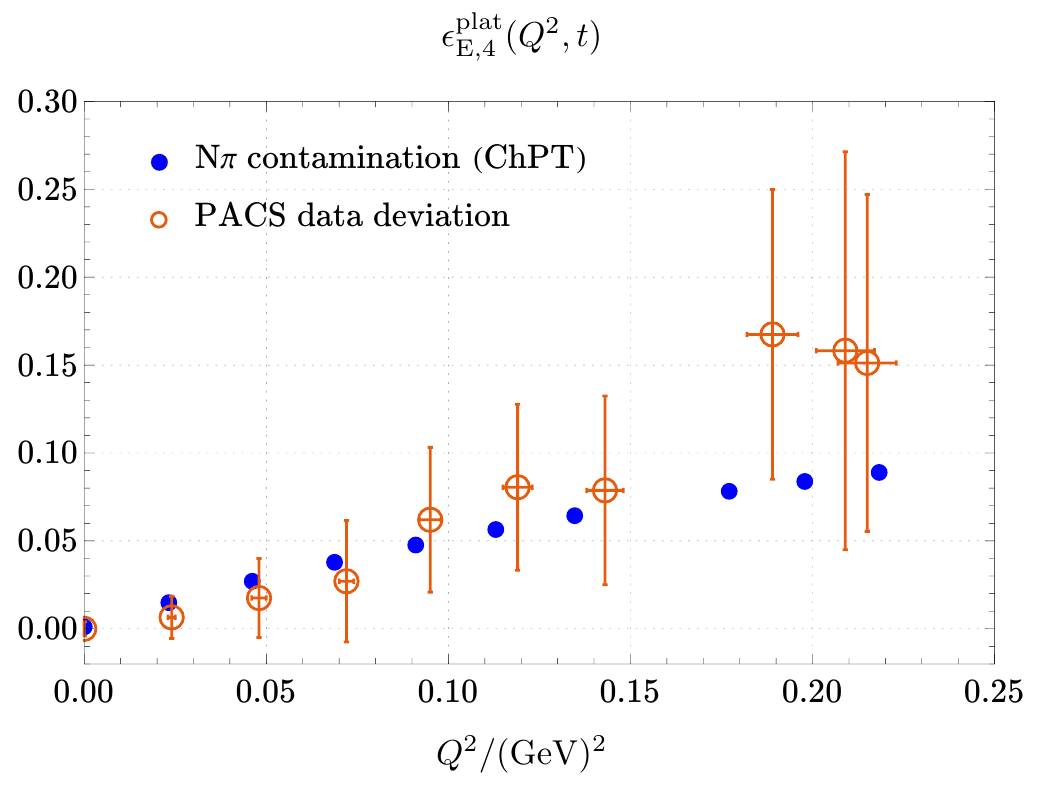}\\[0.6ex]
\includegraphics[scale=1.0]{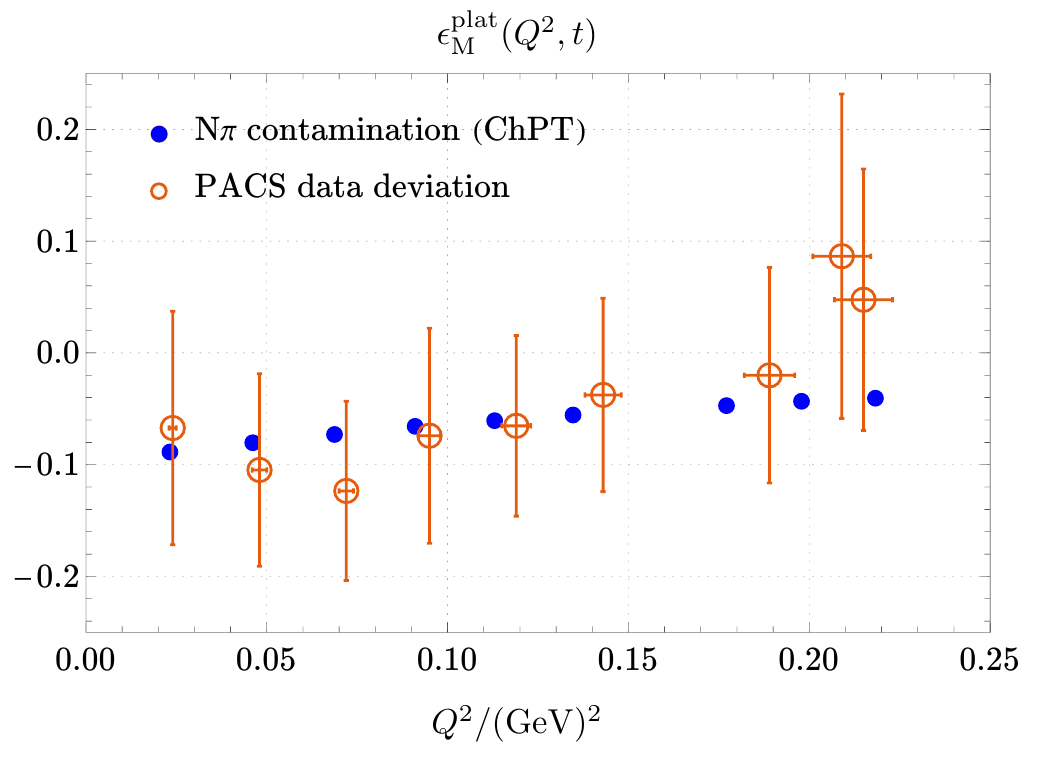}\\[0.6ex]
\caption{\label{fig:PACSGM1dev} The relative deviations $\epsilon^{\rm plat}_{\rm E,4}(Q^2,t)$ (top panel, orange symbols) and $\epsilon^{\rm plat}_{\rm M}(Q^2,t)$ (bottom panel, orange symbols)
of the PACS data \cite{Ishikawa:2018rew} from the Kelly line compared to the ChPT prediction for the deviation due to $N\pi$ states (blue symbols). 
}
\end{center}
\end{figure}

Figure \ref{fig:PACSGM1dev} shows the relative deviation of the PACS lattice data from the experimental results (orange symbols, with error bars) together with the ChPT results for $\eV{4}$ and $\eM$ (blue symbols).\footnote{For the experimental data we use Kelly's parameterization in \cite{Kelly:2004hm}.}  
In both cases the over- and underestimation predicted by ChPT is in good agreement with the lattice data 
and captures qualitatively the $Q^2$ dependence of the deviation. Even though the statistical errors in the lattice data are too large to draw definite conclusions from this comparison, it illustrates that the ChPT predictions for the $N\pi$ state contamination is qualitatively in agreement with what has been observed in lattice QCD data. 

\section{Concluding remarks}

The ChPT results presented here provide an understanding for the anticipated impact of $N\pi$ excited states in lattice computations of the nucleon electromagnetic form factors. ChPT predicts an overestimation of the electric form factor by the plateau or midpoint estimates, and this overestimation gets larger for increasing momentum transfer. For the magnetic form factor we find the opposite, it is underestimated and the smaller the momentum transfer, the larger the underestimation. The size of this effect is about $\pm 5$\% for source-sink separations of 2 fm and momentum transfers smaller than $0.25\,{\rm GeV}^2$. The effect is larger for source-sink separations that are presently accessible with standard simulation methods, for $t\approx1.6\,{\rm fm}$ by roughly a factor two. Therefore, if percent level accuracy is the goal for lattice calculations this source of systematic error is not negligible and needs to be taken care of.

Comparing the results for the $N\pi$-state contamination in the electromagnetic form factors to other results they are comparable in size to the  impact on the axial form factor, but much smaller than the effect in the induced pseudoscalar form factor \cite{Bar:2018xyi}. The origin for this lies in the different symmetry properties of the vector and axial vector currents.  The axial vector current is able to emit (absorb) a single pion that is absorbed (emitted) at the sink (source) of the axialvector 3-pt function, and it is this process that gives rise to a large $N\pi$ contamination in the induced pseudoscalar form factor.
The same process is forbidden for the vector current. Chiral symmetry requires two pions instead of one for the analogous process, and the resulting 3-particle $N\pi\pi$ contamination is expected to be small.

\vspace{4ex}
\noindent {\bf Acknowledgments}
This work was supported by the German Research Foundation (DFG), Grant ID BA 3494/2-1.
\vspace{3ex}


\providecommand{\href}[2]{#2}\begingroup\raggedright\endgroup

\end{document}